\documentclass{anstrans}

\title{Harvesting the Variance Risk Premium in Nuclear and Energy Equities:\\
       A Short-Put Portfolio Derisking Strategy}

\author{Jilang Miao$^{*}$, Nonna Sorokina$^{\dagger}$}

\institute{%
$^{*}$Ken and Mary Alice Lindquist Department of Nuclear Engineering,
Pennsylvania State University, University Park, PA 16802\\
$^{\dagger}$Division of Business, Pennsylvania State University, Dunmore, PA 18512
}

\usepackage{graphicx}
\usepackage{booktabs}
\usepackage{amsmath}
\usepackage{microtype}
\newcommand{\vrpPutEwRet}{8.3}
\newcommand{\vrpPutEwVol}{8.5}
\newcommand{\vrpPutEwSharpe}{0.98}
\newcommand{\vrpPutEwMaxDD}{16.0}

\newcommand{\vrpPutEwCorrStock}{0.62}
\newcommand{\vrpPutIvRet}{7.1}
\newcommand{\vrpPutIvVol}{7.2}
\newcommand{\vrpPutIvSharpe}{0.99}
\newcommand{\vrpPutIvMaxDD}{12.9}

\newcommand{\vrpStockRet}{15.5}
\newcommand{\vrpStockVol}{19.5}
\newcommand{\vrpStockSharpe}{0.80}
\newcommand{\vrpStockMaxDD}{46.1}

\newcommand{\vrpEntries}{6586}
\newcommand{\vrpExcluded}{5}
\newcommand{\vrpTrades}{6581}
\newcommand{\vrpNames}{49}
\newcommand{\vrpMeanEntryDelta}{0.29}

\newcommand{\vrpNameLabels}{41}

\newcommand{\vrpExitDstopPct}{16}

\newcommand{\vrpExitDtePct}{1}

\newcommand{\vrpExitSlPct}{16}

\newcommand{\vrpExitTpPct}{66}
\newcommand{\vrpExitTpAvgPnlPct}{58}
\newcommand{\vrpExitTpHoldDays}{12}

\newcommand{\vrpTrancheSharpeEw}{0.80}

\newcommand{\vrpWinRate}{68}

\newcommand{\vrpThrElevenRet}{7.5}
\newcommand{\vrpThrElevenVol}{10.4}
\newcommand{\vrpThrElevenSharpe}{0.72}
\newcommand{\vrpThrElevenMaxDD}{23.8}
\newcommand{\vrpThrTwelveRet}{5.8}
\newcommand{\vrpThrTwelveVol}{12.0}
\newcommand{\vrpThrTwelveSharpe}{0.49}
\newcommand{\vrpThrTwelveMaxDD}{38.5}
\newcommand{\vrpThrElevenTrades}{2{,}493}
\newcommand{\vrpThrTwelveTrades}{1{,}779}
\newcommand{\vrpPutWorstYear}{2008}
\newcommand{\vrpPutWorstYearRet}{-9.5}

\newcommand{\vrpPutLossYearCount}{2}
\newcommand{\vrpStockRetTwoThousandTwentyFour}{54}

\newcommand{\vrpStockRetInPutWorstYear}{-37.2}
\newcommand{\vrpPutEwWealth}{7.2}
\newcommand{\vrpPutIvWealth}{5.6}
\newcommand{\vrpStockWealth}{29.3}
\newcommand{\vrpRatioMeanTicker}{1.54}
\newcommand{\vrpRatioMeanPooled}{1.45}
\newcommand{\vrpSignalTickersTAboveTwo}{29}
\newcommand{\vrpSignalPosYears}{18}
\newcommand{\vrpSignalTotYears}{25}
\newcommand{\vrpSignalShortfallTwoThousandEight}{-7.2}
\newcommand{\vrpSignalNegYearsList}{2000, 2002, 2007, 2008, 2011, 2018, and 2024}
\newcommand{\vrpSignalNegYearsCount}{7}
\newcommand{\vrpTopOneName}{NRG}
\newcommand{\vrpTopOneT}{12.7}
\newcommand{\vrpTopOneTadj}{2.8}
\newcommand{\vrpTopTwoName}{NXE}
\newcommand{\vrpTopTwoT}{12.1}
\newcommand{\vrpTopTwoTadj}{2.6}
\newcommand{\vrpTopThreeName}{UUUU}
\newcommand{\vrpTopThreeT}{10.0}
\newcommand{\vrpTopThreeTadj}{2.2}
\newcommand{\vrpTopFourName}{MIR}
\newcommand{\vrpTopFourT}{9.6}
\newcommand{\vrpTopFiveName}{LTBR}
\newcommand{\vrpTopFiveT}{9.4}
\newcommand{\vrpTopSixName}{UEC}
\newcommand{\vrpTopSixT}{8.9}

\begin{document}

\maketitle
{\let\thefootnote\relax\footnotetext{$^\S$Corresponding author: Jilang Miao,
\texttt{jlmiao@psu.edu}}}

\section{Introduction}

Nuclear energy plays a growing role in long-term decarbonization plans, yet
private capital flows into nuclear projects remain constrained by the sector's
distinctive risk profile: high up-front capital requirements, decade-scale
construction timelines, and residual regulatory and political uncertainty.
Firms in controversial industries face higher costs of equity \cite{elGhoul2011},
and nearly all nuclear projects in Europe rely on government support due to
limited private capital participation \cite{weibezahn2024}.

Financial engineering offers complementary tools for income-seeking institutional
investors who want nuclear sector exposure without concentrating in raw equity
drawdown risk. Fernandez, Stein and Lo \cite{fernandez2012} showed that a
mega-fund structure can diversify the idiosyncratic risk of early-stage biotech
portfolios, reducing investor uncertainty enough to crowd in private capital.
A sector-level analogue applies to nuclear equities: systematic option writing
harvests elevated implied volatility as recurring income, reducing the effective
risk of holding nuclear positions and potentially broadening the capital base
for the sector over time.

Options markets routinely price implied volatility (IV) above realized volatility
(RV), a wedge known as the variance risk premium (VRP). Selling options provides premium income in exchange for bearing downside
risk below the strike. Bakshi and
Kapadia \cite{bakshi2003} and Carr and Wu \cite{carr2009} document the VRP in
equity index options; Bollerslev, Tauchen and Zhou \cite{bollerslev2009} show
that the VRP forecasts aggregate equity returns. Whether the VRP exists and can
be exploited in individual nuclear and energy equities has not been studied.

We address this question by constructing a systematic short-put portfolio for a
universe of $\sim$45 nuclear and energy-adjacent equities, harvesting the VRP
through cash-secured put writing over 2000--2024. The strategy requires no
complex optimizer and no sentiment signal.
We document (i) statistically significant positive VRP in the majority of names with option coverage,
(ii) a monthly short-put portfolio earning \vrpPutEwRet\% annually at \vrpPutEwVol\% volatility
(Sharpe \vrpPutEwSharpe), versus \vrpStockRet\% at \vrpStockVol\% (Sharpe \vrpStockSharpe) for a passive
equal-weight stock benchmark, with a maximum drawdown of \vrpPutEwMaxDD\% versus \vrpStockMaxDD\%,
and (iii) that exit rules, in particular a delta-based stop,
matter more than entry selection for converting the premium into portfolio returns.
All results are marked to market daily at option mid-prices and stated before transaction costs. The strategy is designed to attract income-seeking private investors
who want premium income that is only partially correlated with equity returns
alongside any residual nuclear equity exposure.

\section{Data and Universe}

We use two WRDS data sources linked by an OptionMetrics--CRSP identifier bridge.
Daily option data come from OptionMetrics IvyDB: strike prices,
best bid/offer, implied volatility, delta, open interest, and expiration date
for all listed U.S. equity options. Daily stock data come from the CRSP Daily
Stock File: prices, returns, and trading volume identified by PERMNO.

Our universe consists of 45 tickers drawn from two independently constructed
datasets: firms identified as nuclear plant operators via NRC plant-ownership
records (utilities with material nuclear generation shares), and constituents of
nuclear-focused ETFs (uranium miners, SMR developers, nuclear service
contractors). Representative names include CEG, NEE, CCJ, UEC, NXE, UUUU,
BWXT, NRG, and SMR. Diversified power producers such as NRG are included
because they held nuclear plant ownership stakes during the sample and their
valuations co-move with nuclear capacity pricing and energy policy.
Coverage spans 2000--2024.
Corporate actions map some tickers to multiple linked securities; \vrpNames{} linked
securities, carrying \vrpNameLabels{} current ticker labels, have sufficient option
coverage and enter the signal construction and backtest.

\section{Strategy Construction}

\subsection{VRP Measurement}

To document the variance risk premium across the universe we estimate realized
volatility with GARCH(1,1) \cite{bollerslev1986}:
\begin{equation}
  \hat{\sigma}^2_t = \omega_i + \alpha\,r_{t-1}^2 + \beta\,\hat{\sigma}^2_{t-1},
  \quad \hat{\sigma}_t = \sqrt{252\,\hat{\sigma}^2_t},
\end{equation}
with $\alpha = 0.09$, $\beta = 0.90$, and $\omega_i$ calibrated per stock
to match each ticker's historical variance: $\omega_i = \overline{r^2}_i(1-\alpha-\beta)$,
an in-sample anchor computed over the full history.
This long-run anchor prevents GARCH from collapsing toward zero during
quiet periods, a flaw of EWMA ($\omega = 0$) that inflates the IV/RV ratio
spuriously.
Parameters $\alpha$ and $\beta$ are fixed at consensus values rather than
estimated by MLE per ticker; per-stock estimation is left for future work.
We compare $\hat{\sigma}_t$ to an ATM implied volatility proxy: the
open-interest-weighted average IV of puts with $|\delta| \in [0.40,\,0.60]$
and DTE $\in [15,\,45]$. The VRP on date $t$ is
$\text{IV}_\text{ATM} - \hat{\sigma}_{t+h}^{\text{realized}}$
where $\hat{\sigma}_{t+h}^{\text{realized}}$ is the realized volatility over the 21 trading
days beginning at $t$.

\subsection{Put Selection}

On the first trading day of each month with a qualifying contract, we select the single put per security that best
matches a target of 0.25-delta and 45-day DTE, scored as:
\begin{equation}
  \text{score} = \frac{\big||\delta| - 0.25\big|}{0.10} + \frac{|\text{DTE} - 45|}{10},
\end{equation}
with eligibility restricted to DTE $\in [35,\,55]$, $|\delta| \in [0.20,\,0.40]$,
open interest $> 10$ contracts, and standard 100-share PM-settled terms with valid quotes.
Single-name chains are sparse: most entry days offer exactly one qualifying contract,
so the realized mean entry delta is \vrpMeanEntryDelta{} rather than the 0.25 target.

\subsection{Exit Rules}

Positions are marked daily and closed at whichever of four conditions occurs first:
(i) premium decays to 50\% of entry mid-price (take-profit),
(ii) mid-price reaches twice the entry premium (stop-loss),
(iii) the put's $|\delta|$ reaches 0.50 (delta stop), or
(iv) DTE $\leq 15$ (early expiration roll).
The delta stop exits positions that have moved substantially against the writer
before the premium stop is hit.
All positions are cash-secured; notional capital per share equals the put strike.
Early assignment is not modeled; since puts are entered out-of-the-money and
exited at 0.50$|\delta|$ at the latest, early exercise is rarely optimal for the holder,
though this remains a modeling simplification.

\section{Results}

\subsection{VRP Evidence}

\begin{figure}[tbp]
  \centering
  \includegraphics[width=\columnwidth]{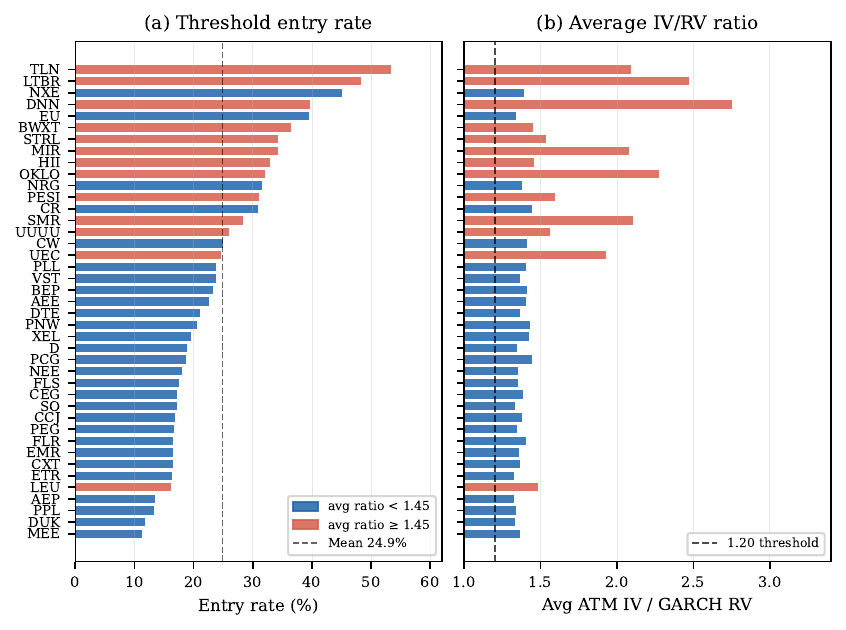}
  \caption{GARCH(1,1) threshold entry rate and average ATM IV / GARCH RV ratio
           by ticker. Red bars: average ratio $\geq 1.45$.}
  \label{fig:signal}
\end{figure}

Figure~\ref{fig:signal} summarizes the ATM IV/RV ratio across the universe.
The mean IV/RV ratio on entry days is \vrpRatioMeanTicker{} averaged equally across tickers
and \vrpRatioMeanPooled{} pooled across all entry-day observations (tickers with more entry
days receive proportionally more weight in the pooled figure). Small
pure-play and SMR names (DNN, LTBR, OKLO, UEC) exhibit the highest ratios;
broadly diversified utilities (SO, AEP, DUK, ETR) the lowest.

\begin{figure}[tbp]
  \centering
  \includegraphics[width=0.9\columnwidth]{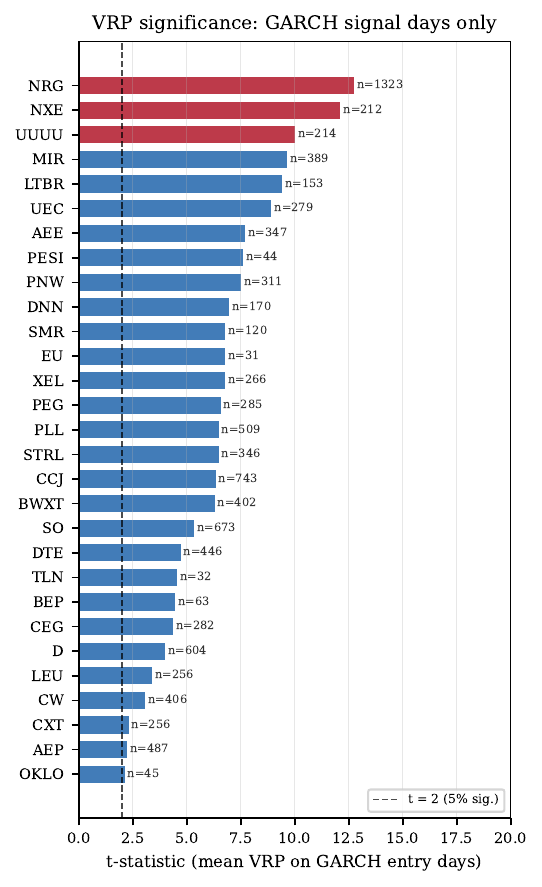}
  \caption{$t$-statistics for mean VRP (IV $-$ forward RV) on entry days.
           Dashed line: $t = 2$ (5\% significance).}
  \label{fig:vrpstat}
\end{figure}

Figure~\ref{fig:vrpstat} shows the $t$-statistics for mean IV $-$ forward RV
on GARCH-signal entry days. \vrpSignalTickersTAboveTwo{} tickers pass $t > 2$ significance.
Nuclear-focused names lead:
\vrpTopOneName{} ($t=\vrpTopOneT$), \vrpTopTwoName{} ($t=\vrpTopTwoT$),
\vrpTopThreeName{} ($t=\vrpTopThreeT$), \vrpTopFourName{} ($t=\vrpTopFourT$),
\vrpTopFiveName{} ($t=\vrpTopFiveT$), \vrpTopSixName{} ($t=\vrpTopSixT$).
These are naive $t$-statistics; 21-day forward RV windows overlap so adjacent
observations share $\sim$21 days of data, overstating effective $n$ up to
$\sim$21-fold. Dividing by $\sqrt{21}\approx 4.6$ as a conservative overlap
adjustment brings the top names to \vrpTopOneName{} ($t\approx \vrpTopOneTadj$),
\vrpTopTwoName{} ($t\approx \vrpTopTwoTadj$), and \vrpTopThreeName{} ($t\approx \vrpTopThreeTadj$).

\begin{figure}[tbp]
  \centering
  \includegraphics[width=\columnwidth]{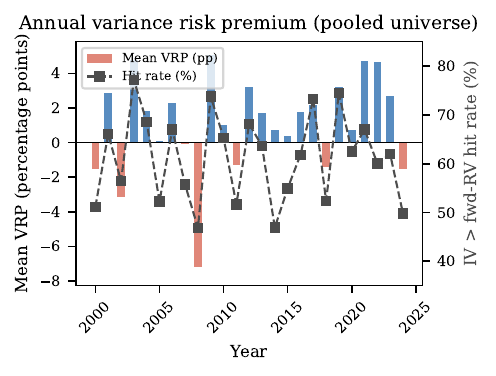}
  \caption{Annual pooled VRP: mean IV $-$ forward RV (bars, left axis) and
           hit rate IV $>$ forward RV (line, right axis).}
  \label{fig:vrpannual}
\end{figure}

The VRP is positive in \vrpSignalPosYears{} of \vrpSignalTotYears{} calendar years (Figure~\ref{fig:vrpannual}).
The \vrpSignalNegYearsCount{} negative years (\vrpSignalNegYearsList) all
correspond to periods of elevated realized volatility that temporarily exceeded
implied levels; 2008 shows the largest shortfall at \vrpSignalShortfallTwoThousandEight{}pp.
The 2024 case is notable: the annual stock return for the universe reached
+\vrpStockRetTwoThousandTwentyFour\%, yet a brief
but sharp volatility spike drove realized volatility above implied levels for
the year as a whole.

\subsection{Portfolio Performance}

We form four portfolio strategies and one benchmark.

\textbf{EW put:} one position per name per month, opened on the first trading day
with a qualifying contract and managed daily by the exit rules above;
equal capital weight across names, and a name's occasional overlapping positions share its capital equally.

\textbf{Inverse-vol put:} same book, with names weighted by the inverse of their GARCH volatility.

\textbf{EW put (threshold 1.1\,/\,1.2):} same book, but a name enters in a month only
on the first trading day whose ATM IV\,/\,GARCH RV ratio clears the threshold;
remaining capital is held in cash.

\textbf{EW stock benchmark:} equal-weight long-only portfolio of the same
universe, rebalanced monthly.

All positions are marked to market daily:
the per-trade daily return is the change in mid-price divided by the strike (cash-secured basis),
aggregated across names and compounded into calendar months.
Of \vrpEntries{} entries, \vrpExcluded{} were excluded because corporate actions during the
holding period made the contract unpriceable with the available data.

\begin{figure}[tbp]
  \centering
  \includegraphics[width=\columnwidth]{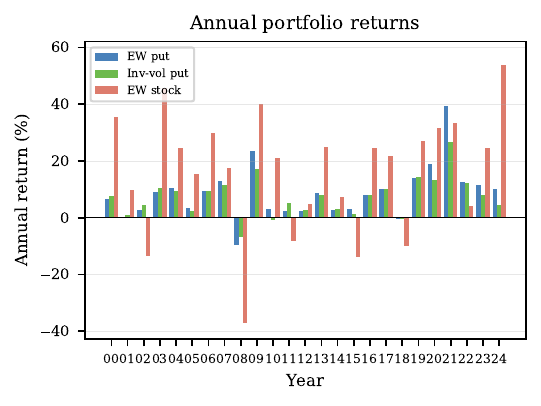}
  \caption{Annual portfolio returns (\%): EW put (blue), inverse-vol put (green),
           and EW stock benchmark (red).
           The put book loses \vrpPutWorstYearRet\% in 2008 versus
           \vrpStockRetInPutWorstYear\% for the stock benchmark.}
  \label{fig:annual}
\end{figure}

Table~\ref{tab:perf} summarizes performance over 2000--2024 (300 months).
The EW put book earns \vrpPutEwRet\% annualized at \vrpPutEwVol\% volatility
(Sharpe \vrpPutEwSharpe), versus \vrpStockRet\% at \vrpStockVol\%
(Sharpe \vrpStockSharpe) for the stock benchmark.
The put book earns less than the stock benchmark but at less than half its volatility,
with the maximum drawdown reduced from \vrpStockMaxDD\% to \vrpPutEwMaxDD\%;
the correlation between the two is \vrpPutEwCorrStock.
Inverse-vol weighting lowers the return slightly and reduces the drawdown to \vrpPutIvMaxDD\%.
The put book loses money in \vrpPutLossYearCount{} of 25 calendar years, worst
\vrpPutWorstYearRet\% in \vrpPutWorstYear{} against \vrpStockRetInPutWorstYear\% for the
stock benchmark that year (Figure~\ref{fig:annual}).

\begin{table}[tbp]
  \caption{Portfolio performance, 2000--2024 (300 months), marked to market daily
           at mid-prices. Cash-secured basis; Sharpe uses zero risk-free rate.}
  \label{tab:perf}
  \centering
  \begin{tabular}{@{}lrrrr@{}}
    \toprule
    Strategy & Ret. & Vol. & Sharpe & MaxDD \\
    \midrule
    EW put               & \vrpPutEwRet\% & \vrpPutEwVol\%  & \vrpPutEwSharpe & \vrpPutEwMaxDD\% \\
    Inverse-vol put      & \vrpPutIvRet\% & \vrpPutIvVol\%  & \vrpPutIvSharpe & \vrpPutIvMaxDD\% \\
    \midrule
    Threshold 1.1        & \vrpThrElevenRet\% & \vrpThrElevenVol\% & \vrpThrElevenSharpe & \vrpThrElevenMaxDD\% \\
    Threshold 1.2        & \vrpThrTwelveRet\% & \vrpThrTwelveVol\% & \vrpThrTwelveSharpe & \vrpThrTwelveMaxDD\% \\
    \midrule
    EW stock bench.      & \vrpStockRet\% & \vrpStockVol\% & \vrpStockSharpe & \vrpStockMaxDD\% \\
    \bottomrule
  \end{tabular}
\end{table}

The risk-adjusted performance depends on the exit rules.
Across the \vrpTrades{} trades, \vrpExitTpPct\% close at the take-profit
(mean \vrpExitTpAvgPnlPct\% of premium in \vrpExitTpHoldDays{} days),
\vrpExitSlPct\% at the premium stop, \vrpExitDstopPct\% at the delta stop,
and \vrpExitDtePct\% at the DTE roll; the overall trade win rate is \vrpWinRate\%.
The threshold variants reduce the trade count substantially
(\vrpThrElevenTrades{} at 1.1; \vrpThrTwelveTrades{} at 1.2) and suffer higher
portfolio volatility 
(\vrpThrElevenVol\% and \vrpThrTwelveVol\% respectively).
Both underperform the unconditional book
(Sharpe \vrpThrElevenSharpe{} and \vrpThrTwelveSharpe{} versus \vrpPutEwSharpe):
the IV/RV ratio that documents the premium does not improve entry timing,
because high-ratio days often precede realized-volatility spikes,
and short puts entered on them lose more through delta than the extra premium recovers.
A NEAR-based sentiment filter \cite{DeriskingNuc} remains a candidate for entry timing.

\begin{figure}[tbp]
  \centering
  \includegraphics[width=\columnwidth]{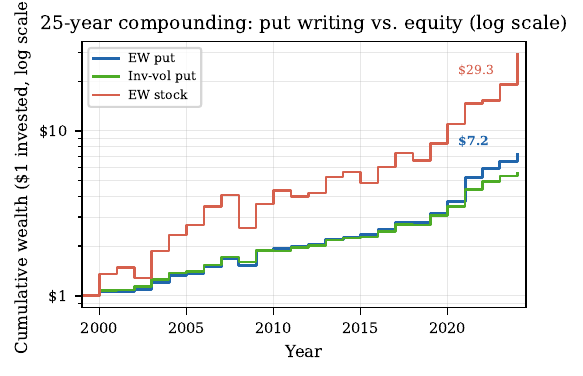}
  \caption{Cumulative wealth of \$1 invested (log scale), 2000--2024.
           EW put (\$\vrpPutEwWealth, blue), inverse-vol put (\$\vrpPutIvWealth, green),
           EW stock benchmark (\$\vrpStockWealth, red). Colors match Figure~\ref{fig:annual}.}
  \label{fig:cumwealth}
\end{figure}

Figure~\ref{fig:cumwealth} shows cumulative wealth of \$1 invested from
January 2000 to December 2024 on a log scale.
The stock benchmark compounds to \$\vrpStockWealth{} with a \vrpStockMaxDD\% maximum drawdown;
the EW put book compounds to \$\vrpPutEwWealth{} with a \vrpPutEwMaxDD\% drawdown.
The put portfolio is better viewed as a lower-risk complement to equity ownership than as a substitute;
a combined allocation retains much of the stock return at a lower drawdown.

\section{Conclusions}
\label{sec:conclusion}

We document a positive and economically significant variance risk premium in nuclear and energy equity options
and show that a simple monthly equal-weight short-put portfolio, disciplined by a delta-based stop,
captures it over 2000--2024.
The strategy earns \vrpPutEwRet\% gross annually on cash-secured capital at
\vrpPutEwVol\% volatility (Sharpe \vrpPutEwSharpe{} versus \vrpStockSharpe{} for a
long-only stock benchmark), with a maximum drawdown of \vrpPutEwMaxDD\% against
\vrpStockMaxDD\%, before transaction costs and the risk-free rate.
The exit rules, rather than an implied-to-realized volatility entry filter, convert the premium
into portfolio returns.

These results suggest that nuclear-focused equity options carry a positive estimated risk
premium that income-seeking investors can harvest systematically, providing a
complement to direct nuclear equity ownership. By offering stable, uncorrelated
cash flows, the strategy may broaden the investor base for nuclear-adjacent
equities, a step toward closing the private financing gap in nuclear energy
development.

Several limitations and extensions bear noting. The backtest uses option
mid-prices throughout; realistic bid-ask execution costs for short puts,
which can be substantial for less liquid names, would reduce reported returns.
The cash-secured capital earns no T-bill return in our model, understating
opportunity cost particularly post-2022; incorporating the risk-free rate
would lower Sharpe ratios. The universe is fixed at today's list of survivors,
introducing survivorship and look-ahead bias; a point-in-time reconstruction
is needed before the performance claims can be treated as investable. Exit
prices assume fills at the first daily mid satisfying the stop or take-profit
condition, which may understate gap risk on adverse overnight moves.
On the model side, GARCH parameters are fixed at consensus values rather than
estimated per ticker.
The design choices were also selected on the full sample, and performance varies across
neighboring variants (entry day of month, eligibility window, take-profit level),
with Sharpe ratios between \vrpTrancheSharpeEw{} and \vrpPutEwSharpe{} and drawdowns
well below the stock benchmark's throughout.

\section*{Acknowledgments}

The authors thank the Pennsylvania State University Institute for Computational
and Data Sciences (ICDS) for computational resources support.
N.~Sorokina acknowledges support from the U.S.\ Department of Energy, Office of
Nuclear Energy, under Award No.\ DE-NE0009381, and from ICDS (RRID: SCR\_025154).

{%
\bibliographystyle{ans}
\bibliography{vrp_nuclear}
}

\end{document}